# Streamlined Pathway (SP) Approach: An Efficient Load Balancer to Enhance Quality of Service

Aymen Hasan Alawadi *

Department of Computer Science, Faculty of Education University of Kufa, 54001 Najaf Governorate, Iraq aymen@uokufa.edu.iq

**Abstract.** Efficient load-balancing mechanisms are critical for maximizing performance and increasing the quality of service (QoS) of data center networks (DCNs). Obtaining the optimal QoS while minimizing resource consumption remains a significant challenge. This paper proposes the streamlined pathway (SP) model, which is a flow scheduling solution that requires minimal statistical knowledge of the DCN data plane. The SP model utilizes the software-defined networks (SDN) paradigm with less information gathered from the DCN data plane besides the traditional hash-based flow scheduling mechanism, the Equal Cost Multi-Path (ECMP). In SDN, the proposed methodology harnesses a minimal, yet powerful set of statistical data extracted from the DCN data plane, including port throughput and elephant flow information on the aggregate switches of the DCN fat-tree topology. Several experiments in addition to theoretical analysis have been conducted to demonstrate the efficiency of the proposed SP model in terms of QoS enhancement. These results confirm that SP outperforms leading techniques such as Sieve, Hedera, and ECMP, concerning bisection bandwidth, DCN link utilization, packet loss, and packet delivery latency.

## 1 Introduction

The unpredicted growing demand for data center (DCN) applications, such as cloud computing, MapReduce, and big data applications, has put substantial pressure on the network infrastructure to deliver the required quality of service (QoS) standards [14]. These applications yield two distinct categories of flows: "mice" flows, characterized by their sensitivity to latency along the path, and "elephant" flows, encompassing long-lived TCP (transmission control protocol) flows that are influenced by available bandwidth along the route. Flows are spread across hosts, using multi-rooted DCN environments to enhance network throughput using various TE strategies. One such widely adopted strategy is the Equal Cost Multipath (ECMP) mechanism [10], which ensures a balanced and random allocation of resources among the multiple paths of the network. Despite its prevalence, ECMP suffers from flow collisions and lacks dynamic adaptability to accommodate varying demands of applications and traffic patterns. Adaptive flow scheduling stands out as the primary solution to elevate QoS standards. To facilitate efficient dynamic flow scheduling and congestion avoidance, it is imperative to conduct thorough monitoring of the DCN environment. This involves gathering metrics from the forwarding devices (i.e., data plane), including port consumption and installed flow numbers, which are essential for making optimal path selections. In the DCN load balancing context, SDN emerged as a robust and efficient paradigm. SDN offers reliable techniques to decouple the control plane, which makes routing decisions, from the data plane, which controls network resources and information in a centralized controller. Generally, the SDN controller connects with the data plane using the southbound application programming interface (API), also known as the OpenFlow protocol. This protocol has parameters called the "flow table", which executes packet lookup and forwarding to the controller [9]. The flow tables are installed as rules to forward the flows through the data plane switches. For new incoming flows, the switch asks the SDN controller for a data plane through a "*packet_in*" message, and the controller processes this request in the controller plane and sends back a "*packet_out*" message, which contains the flow table information. The northbound API is the RESTful interface used for communication between the SDN Controller and other services and applications.

---

* Corresponding author

The proposed method (SP model) introduces an innovative flow scheduling mechanism that effectively employs SDN to use multiple paths in fat-tree DCNs, in conjunction with ECMP hashing. The SP model is based on the multi-commodity flow problem, which is a generalization of the maximum flow problem [12]. The multi-commodity flow problem pertains to identifying distinct flow commodities originating from various sources and heading toward different destinations. However, determining a feasible route for each flow is not a straightforward task [7]. Therefore, this paper introduces the Streamlined Pathway (SP), a novel heuristic and dynamic traffic engineering (TE) approach to addressing the scheduling problem in the DCN environment utilizing a centralized SDN framework. The solution improves the overall bisection bandwidth of the DCN, as well as link utilization, packet loss, and round-robin trip delay (RTT).

The primary contributions of this paper are as follows:

1. **Introduction of Streamlined Pathway (SP) Approach**: Introducing a new and comprehensive DCN load balancing methodology designed to improve QoS and overall enhance network efficiency based on the SDN paradigm.
2. **Empirical validation and performance improvement**: The paper presents empirical results from a comprehensive set of experiments showcasing the effectiveness of the SP approach in boosting DCN performance, thus validating its contributions to the field. The proposed model shows an improvement in the produced bisection bandwidth, link utilization, average packet loss, and packet delivery delay (RTT).

3. **Theoretical analysis and validation of potential performance boundaries**: The paper provides a detailed analysis supported by lemmas and theorems to prove the performance of the proposed model.

## 2 Previous studies

Numerous research studies have introduced solutions to handle flow scheduling and load balancing in DCN that leverage different optimization strategies within the SDN paradigm. Hedera [3] and DevoFlow [6] are some of the most popular methods. Such methods begin by classifying flows as mice and elephant flows. Hedera defines an elephant flow as any flow that exceeds 10% of the DCN link capacity. Then it uses greedy algorithms to determine the best available route for handling each new elephant flow. However, Hedera has been shown to work well only in lightly loaded environments. In contrast, DevoFlow employed a centralized controller to optimize the scheduling of elephant flows while simultaneously addressing mice flow scheduling within the data plane. For both methods, the baseline scheduling algorithm for mice flows is based on the Equal-Cost Multi-Path (ECMP) routing. Where ECMP routes mice through all available paths at the same cost. To minimize the mice flow completion time (FCT), RepFlow was introduced by Xu and Li in [17]. The strategy revolves around replicating the short flows arriving at the host's TCP layer, making use of the DCN multipath routes through ECMP. Implementing such a method involves dedicating two TCP sockets to each host for replication of mice flows. This may require further adjustments to the host's transport layer and coordination between hosts. ECMP has also been exploited in CLOVE [11] for path discovery. In this approach, network flows are organized into set flowlets, defined as groups of packets separated by sufficient intervals. This segmentation enables these flowlets to be efficiently routed along independent paths, mitigating the risk of network congestion. CLOVE leveraged explicit congestion notification (ECN) to forward the new flowlets on uncongested paths. However, while such a mechanism does not require hardware-level modifications. Wang et al. in [16] proposed congestion and quality of service guaranteeing (MCQG), an adaptive load balancing solution based on SDN and ECMP. The solution detects elephant flows on DCN edge switches and reroutes them to another less congested path, considering the link load, the number of elephant flows on a path and transmission delay. In contrast, keeping mice flows handled by ECMP. such a method necessitates a controller overhead associated with continuous flow monitoring. Recent studies employed reinforcement learning algorithms such as Q-learning-based routing optimization in SDN networks [8]. Hassen et al. in [8] proposed congestion avoidance and hybrid exploration methods for multi-path flow-routing problems, to enhance performance metrics like latency, convergence, and computation time. However, key limitations to such studies are their reliance on precise hyperparameter tuning and their scalability challenges in larger, dynamic networks. Contributions extend to addressing the challenges of flow scheduling and load balancing in DCN environments through innovative solutions, which have been presented in Sieve [18] and Oddlab [4]. These approaches leverage the power of SDN coupled with the ECMP routing mechanism to provide effective and scalable solutions to optimize network performance and resource use in DCNs. In Sieve, the main purpose was to reduce mice flows' FCT by finding the ongoing elephant flows on the DCN edges and rerouting part of them to the less busy path. In Oddlab, however, the goal was to implement an efficient load-balancing method that discovers problematic links based on current elephant flows and available bandwidth.

This paper introduces the streamlined path (SP) approach to explore the potential of deploying an effective load-balancing technique that incurs minimal controller overhead. This technique operates seamlessly in symmetric DCN topology, taking advantage of the SDN paradigm without demanding alterations to the network data plane or TCP data packets.

## 3   SP design

The main aspect of SP design is based on the use of proactive paths obtained based on ECMP hashing, coupled with dynamic flow scheduling orchestrated by the central SDN controller. This strategy ensures that each part of the scheduling receives even responsibility in the flow-scheduling operation. Although one might contend that such a mechanism could influence the optimization process, the primary objective is to mitigate the flooding of the *package_in* requests that could overwhelm the SDN controller, thus compromising the efficient determination of optimal paths for incoming flows.

Fig. 1 illustrates the flow chart of the SP model. The model procedure begins by handling the *package_in* requests received from the DCN data plane for a newly arrived flow. Depending on the ECMP hashing function used on the edges, the incoming flow is then processed by the proactive pathways or passed to the SDN controller, which selects the best path. Note that the edge OpenFlow should be configured to evenly distribute the incoming flows between the central controller and data plane OpenFlow switches, using the OpenFlow SELECT feature with two similar bucket weights. Then, the controller collects the required parameters from the data plane including (available bandwidth besides elephant flow information on aggregate switches only). Finally, the SP model will perform the required calculations to build a *package_out* message to install the new flow entries in the data plane.

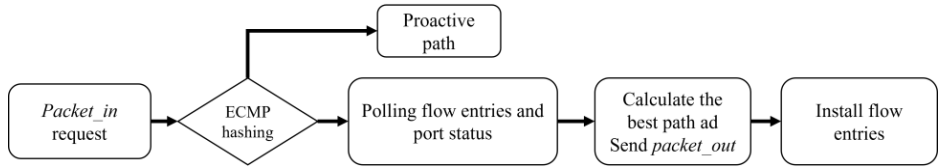

Fig. 1: The flow chart of the SP model

### 3.1 Fat-tree DCN topology

Fat-tree is a multi-rooted topology that is widely used because it is designed to maintain high bisection bandwidth and redundancy across the DCN. As depicted in Fig. 2, the fat-tree has three levels of switches: edge, aggregate, and core switches. These switches are connected in a tree fashion, combining edge and aggregate switches in the pod. In fat-tree topology, each switch is equipped with $K$ ports, and each pod is directly connected to $K^3/4$ hosts. Regarding the port configuration of each switch, it contains $K/2$ ports. Each switch includes two sorts of ports: upstream and downstream. In this configuration, the aggregated switches $K/2$ are linked to $K/2$ upstream ports on the cores, and $K/2$ downstream ports for the edges. Each switch contains two sorts of ports: upstream and downstream. As a result, fat-tree architecture ensures high availability of communication among peers. Such setup provides $(K/2)^2$ equivalent cost paths connecting each the DCN hosts, thereby establishing a robust and fault-tolerant network architecture.

### 3.2 SP architecture

Each DCN edge switch contributes to the flow distribution to the upper $(K/2)^2$ layer switches using SDN-controlled or ECMP methods. With ECMP, a portion of the flows is distributed across multiple ECMP $(K/2)^2$ paths, regardless of the loading status of each path. In contrast, the proposed SDN approach efficiently directs incoming flows along the shortest and least crowded path. In addition, SDN routing dynamically alleviates congestion within the DCN environment by monitoring the presence of ongoing elephant flows, particularly at the upstream end of the aggregates $(K/2)$. It should be noted that all flows in SP have the same priority and that there is no established flow classification on edge switches.

**Lemma 1.** *To optimize the overall flow scheduling, the upstream connections of the aggregate's $K/2$ will detect the maximum number of elephant flows if the load is uniformly distributed on the edges of the DCN $(K/2)$.*

*Proof.* Assume a fat-tree DCN shown in Fig. 2 with uniform traffic patterns, with N edge switches and M upstream link of aggregate switches. Each edge switch has multiple paths to multiple core switches, typically $(K/2)^2$ paths. Incoming flows are distributed equally (i.e., do not consider flow size, leading to suboptimal bandwidth distribution) in the ECMP-based method by each edge switch $E_i$ ($i \in \{1, 2, \ldots, N\}$) across all its $P$ paths to the aggregate switches. This indicates that the average load of each path from the edge switch is equal to 1.

$$\text{Each edge switch } E_i = \frac{\text{Total load from } E_i}{P} \quad (1)$$

As a result, the average load on each upstream link of an aggregate switch $A_j$ ($j \in \{1, 2, \ldots, M\}$) will be the sum of the equally distributed loads of each

edge switch (Equation 2)

$$\text{Load on aggregate switch } A_j = \sum_{i=1}^{N} \left( \frac{\text{Load from } E_i \text{ to } A_j}{P} \right) \quad (2)$$

Hence, elephant flows are distributed across multiple upstream links rather than concentrated on a single link, thus optimizing overall flow scheduling and reducing the likelihood of congestion. Therefore, each link between aggregate and the core switches can sense a portion of the elephant flows. Such real-time load information optimizes the scheduling algorithms (e.g., SDN controller) so the SDN controller can operate more efficiently because the number of *packet_in* requests will be minimized since the controller's decisions are spread more uniformly throughout the network, reducing computational overhead.

Fig. 2 illustrates the DCN topology of a K fat-tree. This configuration helps to demonstrate the load distribution from the edges to both aggregate and core switches, supporting Lemma 1 that the equal load distribution from the edge switches maximizes the detection of elephant flows and optimizes the overall flow scheduling.

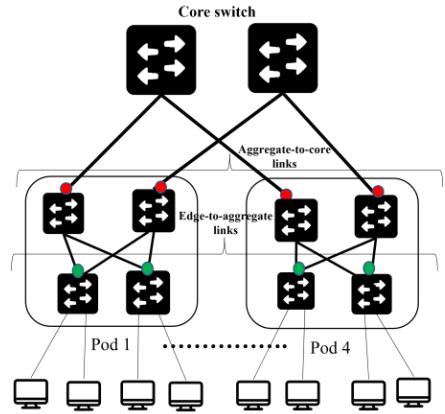

Fig. 2: Fat-tree DCN: Network load traffic on edge-to-aggregate links (green) and elephant flow identification on aggregate-to-core switches (red)

**Lemma 2.** *Scheduling incoming flows along paths with the highest available bandwidth and the fewest existing elephant flows on aggregate switches minimizes network congestion, improving throughput, and reducing latency with manageable computational overhead.*

*Proof.* Let $F_i$ be an incoming flow, $R(P_j)$ represents the path free available bandwidth $P_j$, and $E(P_j)$ indicates the ongoing elephant flows discoverer on the path $P_j$. Define $\alpha$ as a constant that balances the trade-off between the free available bandwidth and the ongoing elephant flow number.

To select the optimal path ($P_{\text{optimal}}$), the path selection process should maximize the following objective function:

$$P_{\text{optimal}} = \arg\max_{P_j} \left( R(P_j) - \alpha \cdot E(P_j) \right) \quad (3)$$

This process will result in paths with more accessible bandwidth while ranking paths with more active elephant flows lower. Hence, such flow scheduling will ensure a more balanced network and help prevent potential bottlenecks without requiring flow rescheduling. Now, by scheduling flows to the optimal chosen path, the chances of flow congestion will be minimized, leading to higher throughput for the entire network (i.e., the amount of data transferred per unit of time (e.g., Mbps)). Therefore, the aggregate throughput $T_{\text{aggregate}}$ is maximized as shown in Equation 4:

$$T_{\text{aggregate}} = \sum_{i=1}^{N} \left( \frac{F_i}{P_{\text{optimal}}} \right) \tag{4}$$

Path latency $L\_path$ can be approximated as the time required to transmit a unit of data. As a result of the minimized congestion, queueing delays on each path are reduced, leading to lower overall latency (Equation 5).

$$L_{\text{path}} \approx \frac{1}{T_{\text{aggregate}}} \tag{5}$$

Note that network latency depends on various factors including queueing delay, flow propagation delay, transmission delay, and processing delay. Equation 5 is used as an approximation in this paper, assuming minimal queuing delay. Analysis of the computational and memory complexity of the proposed model for feature collection from switches in a DCN will be conducted using the following metrics: **Available Bandwidth for All Ports**: Let $N_s$ be the total number of switches in the DCN, Let $P_r$ be the number of ports per switch, for each port on each switch, the available bandwidth must be collected, and computational Complexity is $O(N_s \times P_r)$. **Number of ongoing Elephant Flows on Aggregate Upstream Switches**: Let $A$ be the number of aggregate upstream switches, and Computational complexity is $O(A)$. Note that the number of aggregate upstream switches $A$ is typically much smaller ($\frac{K^3}{4}$) than the total number of ports $N_s \times P_r$, which is ($\frac{5K^3}{4}$). As a result, the overall computational complexity will be dominated by the available bandwidth collection process, so the computational complexity of the model will be $\approx O(N_s \times P_r)$.

The total memory needed to reserve the bandwidth that is accessible and the number of ongoing elephant flows (Total Memory Complexity = $O(N_s \times P_r) + O(A)$) is the memory complexity of the suggested model. Since the total number of aggregate upstream switches is much smaller than the total number of ports, the total memory complexity will be simplified to $\approx O(N_s \times P_r)$. Such complexities highlight the feasibility of implementing the proposed model in large-scale DCNs without overwhelming the SDN controller, thus optimizing overall network performance and load balancing.

The proof in Lemma 2 shows that the proposed model will handle the incoming flows efficiently and resolve the network congestion with less cost if such flows are distributed to the SDN controller.

Algorithm 1 describes the detailed process for finding the optimal path for the flows handled by the SDN controller. Note that elephant flow detection depends on port flow rate consumption with a byte count of 50 kbps as recommended by Roy et al. in [13].

---

**Algorithm 1:** SP adaptive flow scheduling algorithm.

---

**Input:** SP (scrip, dstip, min_bw, max_fno)
**Output:** The best path
**Step 1: Sort paths by shortest paths (minimum number of switches)**
**foreach** *path i from 1 to k* **do**
    **if** *shortest_path[i] = shortest_p(scrip, dstip)* **then**
        Save all available shortest paths in *shortest_p* list;
**Step 2: Sort paths by minimum number of elephant flows on aggregate switches**
**foreach** *path j in shortest_p* **do**
    *fno_paths(scrip, dstip)* ← *fno_ag_up*;
    *min_fno_path* ← *min(fno_paths(scrip, dstip, fno_ag_up))*;
    Sort and save paths in *min_fno_path* list;
**Step 3: Filter paths by maximum available bandwidth**
**foreach** *path w in min_fno_path* **do**
    *bw_paths(scrip, dstip)* ← *path_bw*;
    *max_bw_path* ← *max(bw_paths(scrip, dstip, path_bw))*;
    Sort and save paths in *max_bw_path* list;
**return** The first path in *max_bw_path* as the best path;

---

## 4 Theoretical analysis

A theorem based on Lemmas 1 and 2 is proposed to construct a theoretical principle governing the SP model's adaptive flow scheduling. The proposed theorem provides a foundation for designing scheduling algorithms that improve throughput and reduce latency while addressing computational and memory complexities.

**Theorem 1.** *In a fat-tree DCN with k pods and N switches, the optimal flow scheduling and load distribution are achieved*

*if: (1) The load is evenly distributed across the k/2 edge switches, and (2) Each incoming flow f is scheduled to the current optimal path (P_optimal).*

*The network then achieves globally optimal flow scheduling, which is characterized by the following: Maximized detection and balanced distribution of elephant flows across aggregation switches, minimized overall network congestion, maximized network throughput, reduced end-to-end latency, improved utilization of network resources, and scalable performance with reasonable computational complexity.*

### 4.1   Performance bounds

An investigation is carried out to determine the performance limitations of the proposed lemma by studying the distribution of elephant flows and their consequent impact on network performance, specifically throughput, latency, and load balancer efficiency. Understanding the performance restrictions can help identify the primary parameters influencing the proposed technique's success and provide information for further optimization.

## 4.2 Throughput

**Upper Bound:** The SP adaptive mode produces the maximum achievable throughput ($T_{max}$), which occurs when the load is evenly distributed across all available paths $P$ between the source and destination. It can be expressed in Equation 6, where $E_i$ represents each edge switch ($K/2$):

$$T_{max} = \sum_{i=1}^{k/2} \frac{\text{Total load from } E_i}{P} \tag{6}$$

**Lower Bound:** Conversely, when the load is distributed unevenly, certain links may become congested, resulting in a decreased minimum yielded throughput $T_{min}$ as given by Equation 7:

$$T_{min} = \min \frac{\text{Total load from } E_i}{P} \tag{7}$$

## 4.3 Latency

**Upper Bound:** The minimum latency achieved with optimal load distributions is denoted by $L_{max}$, which corresponds to the scenario in which the network operates at maximum throughput with minimal queuing delays and is given by Equation 8:

$$L_{max} \approx \frac{1}{T_{max}} \tag{8}$$

**Lower Bound:** The network may experience significant queuing delays in the case of uneven load distribution, which produces a higher minimum latency, denoted as $L_{min}$, and calculated as Equation 9:

$$L_{min} \approx \frac{1}{T_{min}} \tag{9}$$

## 4.4 Load balancing efficiency

**Upper Bound:** Load balancing reflects the degree to which the load is distributed throughout the network fabrics such that it reaches its maximum value $E_{max}$ when the load is perfectly balanced. This can be expressed as in Equation 10:

$$E_{max} = 1 - \frac{\sum_{j=1}^{k/2} \left( \frac{L_{A_j}}{\sum_{i=1}^{k/2} L_{E_i}} - \frac{1}{P} \right)^2}{k/2} \tag{10}$$

Where $L_{A_j}$ is the load on the aggregate links $j$, $L_{E_i}$ is the total load of all the edge links, and $k/2$ represents the number of aggregate switches or paths in a single pod of the network.

The variance is estimated based on the ideal distribution and normalized by the number of available paths. The sum in the equation measures the deviation of the load ratio on each link from the ideal value $\frac{1}{P}$. Hence, the smaller the deviation value, the closer it is to ideal balancing, thus maximizing $E_{max}$. **Lower Bound:** The efficiency of load balancing is dropped to 0 when the load is distributed unevenly, denoted as ($E_{min}$ = 0).

The above bounds emphasize the importance of the adaptive SP load distribution in optimizing network performance. By ensuring that the load from the edge switches is evenly distributed, the proposed scheduling approach can maximize throughput, minimize latency, and achieve efficient load balance.

## 5 Empirical performance evaluation

In this section, the performance evaluation of the results of the proposed model is discussed and compared with other methods, including adaptive methods such as Sieve [18], Hedera [3], in addition to the static hashing method (ECMP). Furthermore, the SP performance is compared with an ideal network case in which all hosts are directly connected to a single non-blocking switch. The methods used to perform the experiments and generate the required traffic are then explained.

### 5.1 Methodology

SP was implemented based on previous studies (Sieve [18] and the Oddlab link failure awareness model [4]). The proposed model was developed and operated on a Ryu controller [15] using mininet [2] to create the DCN environment ($K-4$ fat-tree). In topology simulation, the link capacity was set at 10 Mbps to establish a 1:1 bisection bandwidth scenario (when the DCN is divided into two halves [5]), ensuring an equal load distribution throughout the DCN environment. Traffic is generated synthetically using tools such as Iperf [1] to simulate elephant TCP (transmission control protocol) flows in addition to the ping command for ICMP (control message protocol) mice flows. A random traffic pattern simulates realistic traffic transformation across a DCN bisection to ensure that the host and destination are chosen randomly.

**5.2 Evaluation metrics**

In this study, the efficiency of the proposed model was evaluated using various metrics that affect the operation of the DCN to provide the required level of service, such as (aggregate bisection bandwidth, network link utilization, delivery packet loss, as well as packet delay (round-trip delay time (RTT)).

# 6 Results and Discussion

Fig. 3a shows the average bisection bandwidth derived using DCN flow scheduling methods. Except for the non-blocking (ideal situation) in which the hosts were linked directly to the main switch, the SP model exhibited the maximum bisection bandwidth among the investigated approaches. The SP model optimizes at 15% compared to Sieve and 14% for Hedera and ECMP. This is because the SP model splits incoming flows more efficiently among the available paths with fewer elephant flows, thus avoiding congested routes; however, alternative methods produced nearly identical results.

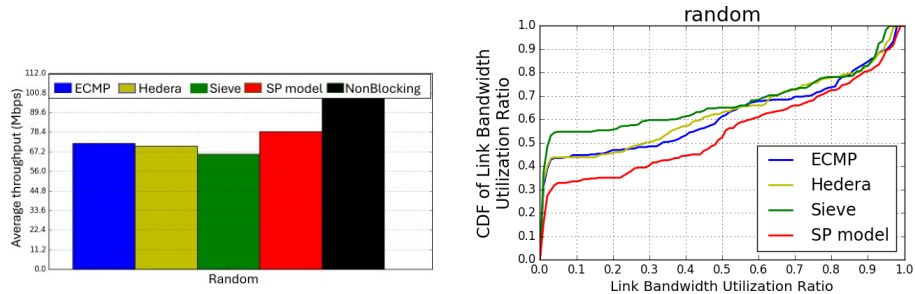

(a) Average bisection bandwidth benchmark results of the compared methods.

(b) The results of comparing the link utilization ratios of the different methods.

Fig. 3: Comparison of throughput and link utilization metrics for different methods.

In terms of the ink usage ratio of the DCN, Fig. 3b shows the cumulative distribution function (CDF) of the utilization ratio of the bandwidth of the link relative to the number of DCN links used. The results demonstrate that the SP model evenly distributed the loads across all potential links. To better understand the results, a CDF ratio of 0.5 was considered to measure the bandwidth consumption for 50% of the DCN links. The results demonstrate that comparable approaches are achieved (48%,33%, 28%, and 2%) for SP, Sieve, Hedera, and ECMP, respectively. At the end of the test, the results for the 90[th] percentile (90% in CDF) ended up in the consumption ratios (96%, 94%, 92%, and 94%) for SP, Sieve, Hedera, and ECMP. These findings confirm that SP extensively utilizes its available bandwidth. As a result, this optimizes DCN resources

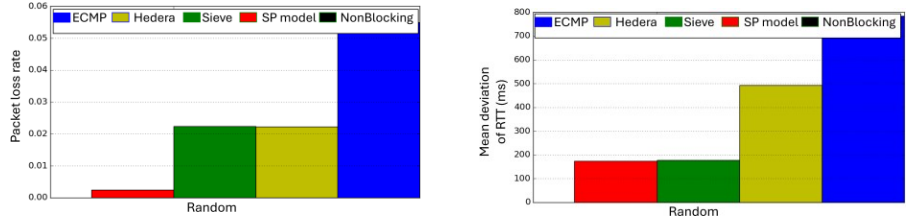

(a) Average packet loss rates for the mice flow results of the compared methods.

(b) The mean deviation results of RTT values for mice flows.

Fig. 4: Comparison of packet loss rates and RTT deviations for mice flows using different methods.

to maximize throughput. Furthermore, these results suggest that Hedera and ECMP did not fully utilize DCN connections. Severe under-use slows data flow, particularly when high throughput is necessary for DCN applications. A further test was conducted to prove that the results of SP utilization did not affect the productivity of the mice flows. This test includes a stream of packets generated by ICMP ping along with elephant flows generated among the DCN hosts. Fig. 4a confirms that the packet loss ratio of the SP model is significantly lower than that of the other methods. Although the loss rates for both Sieve and Hedera were similar as a result of elephant flow rerouting, there was a high packet loss in the ECMP case due to recurrent flow collision. To illustrate the delay of each arrived packet, Fig. 4b reveals that the SP model also has a lower mean deviation for the RTT values, followed by Sieve, Hedera, and ECMP. These findings further confirm that the SP model not only achieves fewer packet losses but also exhibits reduced latency for each arriving packet.

The benchmark results demonstrate that the proposed SP model outperforms other methods in terms of different DCN performance metrics. This is due to efficient flow scheduling among all available paths between sources and destinations. Consequently, SP presents a scalable yet efficient DCN flow scheduling approach that requires no data plane modification.

## 7   Limitations and potential future research directions

Although the efficiency of the proposed model has been proven theoretically and empirically. It is also crucial to consider some potential limitations and explore opportunities for future research to enhance their functionalities. One of the key limitations of the model is that the performance evaluation and theoretical analysis were conducted primarily on the DCN topology of fat-trees. Therefore, the adaptability and functionality of the model in further topologies, such as Clos networks, Dragonfly, or Jellyfish, remain to be investigated. However, the SP model is efficient in detecting elephant flows using a threshold-based approach and lacks sufficient precision for early differentiation between mice and elephant flows. Hence, incorporating an advanced flow classification approach, such as machine learning-based models, can improve the classification decision of scheduling operations.

## 8 Conclusions

This paper presents the streamlined pathway (SP) model, which is a new DCN flow scheduling approach that combines proactive equal-cost multipath (ECMP) routing with adaptive software-defined networking (SDN)-based control to enhance network performance. The primary contributions of the SP approach are multifaceted. First, to reduce the amount of network information acquired and minimize the overhead of the SDN controller, only minimal statistical insights from the DCN data plane were considered, including port throughput in addition to elephant flow information and aggregate switches of the fat-tree DCN topology. Second, the theoretical analysis, which includes lemmas and theorems, establishes the foundation of the SP approach. Third, a comprehensive empirical evaluation to validate the SP approach. These results demonstrate that the SP model outperforms existing methods such as Sieve, Hedera, and ECMP in terms of bisection bandwidth, network link utilization, delivery packet loss, and RTT latency. Although the SP approach is efficient, this study also discusses potential limitations and investigates future research directions that can be explored to further optimize an advanced solution for load balancing in DCNs.